\documentclass[12pt,letter]{article}

\usepackage{graphicx, epsfig, color}
\usepackage{amssymb,amsmath}

\def\eslt{E_T^{\rm miss}}
\def\delew{\Delta_{\rm EW}}

\def\delhs{\Delta_{\rm HS}}
\def\delbg{\Delta_{\rm BG}}
\def\to{\rightarrow}

\def\bi{\begin{itemize}}
\def\ei{\end{itemize}}

\def\tst{\tilde t}

\def\tg{\tilde g}

\def\tw{\widetilde W}
\def\tz{\widetilde Z}
\def\alt{\lesssim}

\def\be{\begin{equation}}  
\def\ee{\end{equation}}  
\def\bea{\begin{eqnarray}}  
\def\eea{\end{eqnarray}}  
\def\beas{\begin{eqnarray*}}  
\def\eeas{\end{eqnarray*}}  
\newcommand\prd[3]{{\it Phys.\ Rev.\ }{\bf D #1} (#2) #3}

\newcommand\plb[3]{{\it Phys.\ Lett.\ }{\bf B #1} (#2) #3}
\newcommand\jhep[3]{{\it J. High Energy Phys.\ }{\bf #1} (#2) #3}

\newcommand\npb[3]{{\it Nucl.\ Phys.\ }{\bf B #1} (#2) #3}

\newcommand{\hepph}[1]{hep-ph/#1}

\begin{document}
\begin{titlepage}
\begin{flushright}
CERN-PH-TH/2013-317 \\
UH-511-1227-14
\end{flushright}

\vspace{0.5cm}
\begin{center}
{\Large \bf  Monojets and mono-photons
from light higgsino pair production at LHC14
}\\ 
\vspace{1.2cm} \renewcommand{\thefootnote}{\fnsymbol{footnote}}
{\large Howard Baer$^1$\footnote[1]{Email: baer@nhn.ou.edu },
Azar Mustafayev$^{2,3}$\footnote[2]{Email: azar@phys.hawaii.edu } and
Xerxes Tata$^2$\footnote[4]{Email: tata@phys.hawaii.edu } 
}\\ 
\vspace{1.2cm} \renewcommand{\thefootnote}{\arabic{footnote}}
{\it 
$^1$Dept. of Physics and Astronomy,
University of Oklahoma, Norman, OK 73019, USA \\
}
{\it 
$^2$Dept. of Physics and Astronomy,
University of Hawaii, Honolulu, HI 96822, USA \\
}
{\it 
$^3$Theory Division, CERN, CH-1211 Geneva 23, Switzerland\\
}
\end{center}

\vspace{0.5cm}
\begin{abstract}
\noindent 
Naturalness arguments imply the existence of higgsinos lighter than
200-300~GeV. However, because these higgsinos are nearly mass
degenerate, they release very little visible energy in their decays, and
signals from electroweak higgsino pair production typically remain
buried under Standard Model backgrounds.  Moreover, gluinos, squarks and
winos may plausibly lie beyond the reach of the LHC14, so that signals
from naturalness-inspired supersymmetric models may well remain hidden
via conventional searches.  We examine instead prospects for detecting
higgsino pair production via monojets or mono-photons from initial state
radiation.  We find typical signal-to-background rates at best at the
1\% level and without any spectral distortions, leading to rather
pessimistic conclusions regarding detectability via these channels.

\vspace*{0.8cm}

\end{abstract}

\end{titlepage}

\section{Introduction}


The minimization of the (renormalization group improved one-loop)
electroweak scalar potential of the Minimal Supersymmetric Standard
Model (MSSM) leads to the well-known relation \cite{wss},
\be 
\frac{M_Z^2}{2} = \frac{m_{H_d}^2 + \Sigma_d^d - 
(m_{H_u}^2+\Sigma_u^u)\tan^2\beta}{\tan^2\beta -1} -\mu^2 
\label{eq:mz}
\ee 
where the running potential parameters are evaluated at the scale
$M_{\rm SUSY}=\sqrt{m_{\tst_1}m_{\tst_2}}$
and where $\Sigma_u^u$ and $\Sigma_d^d$ are radiative corrections that
arise from the derivatives of $\Delta V$ evaluated at the potential minimum.
The sensitivity of $M_Z^2$ to the input parameters has been used
to construct the necessary (though not sufficient) condition for naturalness
defined by the {\it electroweak fine-tuning measure} \cite{ltr,rns},
\be 
\delew \equiv max_i \left|C_i\right|/(M_Z^2/2)\;, 
\ee 
where $C_{H_d}=m_{H_d}^2/(\tan^2\beta -1)$,
$C_{H_u}=-m_{H_u}^2\tan^2\beta /(\tan^2\beta -1)$ and $C_\mu =-\mu^2$.
Also, $C_{\Sigma_u^u(k)} =-\Sigma_u^u(k)\tan^2\beta /(\tan^2\beta -1)$
and $C_{\Sigma_d^d(k)}=\Sigma_d^d(k)/(\tan^2\beta -1)$, where $k$ labels
the various loop contributions included in Eq.~(\ref{eq:mz}).
Expressions for the $\Sigma_u^u$ and $\Sigma_d^d$ are given in the
Appendix of the second paper of Ref.~\cite{rns}. 

Note that $\delew$ is essentially determined by the SUSY spectrum. It is
independent of both the underlying mechanism by which the
super-partners acquire their masses and of the messenger scale --
$\Lambda$ -- at which this mechanism is operative. 
This is in sharp contrast to conventional measures of fine-tuning 
such as $\delbg$ \cite{ellis,bg} or $\delhs$ \cite{kn,papucci,rns} 
where corrections such as 
$\sim m_{H_u}^2(\Lambda)\ln\left(\frac{\Lambda^2}{m_{\rm SUSY}^2}\right)$ 
lead to very high values of these fine-tuning measures 
especially in models -- such as mSUGRA -- 
where the parameters defined at a very high energy scale. 
There is, of course, no contradiction since small $\delew$ is, 
as we have mentioned, just a necessary condition for 
fine-tuning~\cite{rns,comp}.

An immediate consequence of Eq.~(\ref{eq:mz}) is that models with 
values of $\mu^2 \gg M_Z^2$ are necessarily fine-tuned. We emphasize
that although we have used the electroweak scale minimization conditions to
argue this, the same conclusion follows even with the use of 
popular fine-tuning measures. 
This is because $\mu^2$ runs very little between
$M_{\rm GUT}$ and $M_{\rm SUSY}$ so that the
sensitivity of  $M_Z^2$ to $\mu_0$, the GUT scale value of $\mu$, is
changed by just $\sim 10$\%.\footnote{This simple
  fact often remains obscured because the values of both $\delhs$ and 
$\delbg$ are defined by the input parameter that $M_Z^2$ is most
sensitive to, and this is almost never $\mu_0^2$.} 
We thus conclude that a small value of $\mu^2/(M_Z^2/2)$ is 
{\em a robust and necessary condition for naturalness irrespective 
of the fine-tuning measure   that is used.} 
Stated differently,  models with higgsinos heavier than
200~GeV (300~GeV) necessarily have a fine-tuning worse than 10\% (3\%). 
Experimental probes of light higgsinos pair production can thus decisively 
probe naturalness of SUSY models.\footnote{Note that the link between 
fine-tuning and the higgsino mass breaks down if the dominant contribution 
to the higgsino mass is non-supersymmetric \cite{sundrum}. 
If there are no singlets that couple to higgsinos, 
such a contribution would be soft. 
However, in all high scale models that we know, the higgsino mass has
a supersymmetric origin.}

Motivated by these considerations, we have examined the
spectra and aspects of the phenomenology that result in models where
$\delew\sim 10-30$. Typically, the dominant radiative corrections to
Eq.~(\ref{eq:mz}) come from the top-squark contributions
$\Sigma_u^u(\tst_{1,2})$.  For negative values of the trilinear soft
term $A_t$ somewhat larger than the GUT scale scalar masses, each of
$\Sigma_u^u(\tst_{1})$ and $\Sigma_u^u(\tst_{2})$ can be minimized
whilst lifting up $m_h$ into the 125~GeV regime \cite{ltr} as required by
the discovery of the Higgs boson at the LHC~\cite{atlas_h,cms_h}.
Upon requiring no large independent contributions to Eq.~(\ref{eq:mz})
(which would necessitate fine-tuning of the remaining parameters to keep
$M_Z$ at $\simeq 91.2$~GeV), we find that 
\bi
\item $|\mu| \sim 100-300$~GeV (the closer to $M_Z$ the better),
\item $m_{H_u}^2$ is driven radiatively to only small negative values,
\item the top squarks which enter the $\Sigma_u^u$ radiative corrections are 
highly mixed and lie at or around the few~TeV scale and
\item in order to keep $m_{\tst_{1,2}}$ from growing too large, 
the gluino mass is also bounded from
above, in this case by $m_{\tg}\alt 4-5$~TeV. 
\ei

Sparticle mass spectra consistent with low $\delew$ can readily yield a
value of $m_h\sim 125$~GeV whilst evading LHC8 search limits on squarks,
gluinos and top-squarks \cite{atlas_susy,cms_susy}, and at the same time
maintaining low electroweak fine-tuning, our necessary condition for
naturalness.  The key feature of the mass spectra implied by
Eq.~(\ref{eq:mz}) is the existence of four light higgsinos --
$\tw_1^\pm$, $\tz_1$ and $\tz_2$ -- all with mass $\sim |\mu| \sim
100-300$~GeV. While these light higgsinos can be produced at LHC at
large rates, their compressed spectra with mass gaps
$m_{\tz_2}-m_{\tz_1}\sim m_{\tw_1}-m_{\tz_1}\sim 10-30$~GeV results in
only soft visible energy release from their three-body decays; this makes
signal extraction from SM background exceedingly difficult, if not
impossible.

A new signature endemic to models with light higgsinos has also been
pointed out in Ref.~\cite{lhcltr}: $pp\to \tw_2^\pm\tz_4\to (\tz_{1,2}
W^\pm)+(\tw_1^\mp W^\pm)$ which results in hadronically quiet -- because
the decay products of $\tw_1$ and $\tz_2$ are soft -- same sign diboson
events (SSdB). The 300~fb$^{-1}$ LHC14 reach for SSdBs extends to a
wino mass of about 700~GeV. This corresponds to $m_{\tg}\sim 2.1$~TeV in
models with gaugino mass unification, somewhat larger than the LHC14 reach
for gluino pair production \cite{rnslhc}. 
Confirmatory signals will also be present in multilepton 
channels~\cite{rnslhc}. 
Since $m_{\tg}$ can extend up to $4-5$~TeV while maintaining low 
$\delew \alt 30$, then
LHC14 can probe only a fraction of the parameter space of natural SUSY 
in this manner.


An alternative LHC search strategy has been proposed in a variety of
papers (for a summary and detailed references, see {\it e.g.}
Ref.~\cite{matchev}), namely to look for initial state QED/QCD-radiation
off WIMP pair production.  Much of this
work~\cite{zhang,beltram,goodman,rajaraman,fox,cms,atlas} has been
carried out using effective operator analyses. Here, it is assumed that
the interactions between the dark matter particle and SM quarks occur
via very heavy mediators (usually $t$- and $u$-channel squarks in the
context of the MSSM bino-like WIMP) so that the contact approximation is
valid. It is clear that for MSSM higgsino pair production the contact
interaction approximation breaks down very badly because higgsinos are
dominantly produced by collisions of quarks and anti-quarks (inside the
protons) via $s$-channel $Z$ exchange. Since higgsinos are necessarily
heavier than $\sim 100$~GeV, the $Z$-boson propagator suppresses the
amplitude for higgsino production by an extra factor of $\hat{s}$
relative to the contact-interaction approximation.  This results in a
suppression of the cross-section where the higgsino pair is produced
with large invariant mass. Since radiation of hard gluons or photons is
most likely in this regime, the contact interaction approximation will
badly overestimate the cross section for high $E_T$ monojet and
mono-photon events, as has already been emphasized in
Ref.~\cite{buchmueller}. As a result, constraints \cite{cms,atlas} using
effective operator analyses, therefore, do not apply in the case that
the light SUSY states are higgsinos.

In a recent analysis, Han {\it et al.}~\cite{han} have computed the
monojet signal in the natural SUSY framework with light higgsinos using
the complete matrix element.  An advantage of applying this technique to
models with light higgsinos is that one is not restricted to just WIMP
($\tz_1$) pair production, but one may radiate off gluons or photons in
several other reactions as well: $pp\to \tw_1^+\tw_1^-$, $\tz_1\tz_2$,
$\tz_2\tz_1$ and $\tw_1\tz_{1,2}$, since again the heavier higgsino
decay debris is expected to be soft (unless highly boosted) at LHC.
Including all the relevant contributions, these authors claim that
LHC14, with an integrated luminosity of 1500~fb$^{-1}$ will be able to
probe higgsinos up to 200~GeV at 5$\sigma$~\cite{han}. If their results
hold up to scrutiny, it will imply that experiments at the high
luminosity upgrade of the LHC will decisively probe SUSY models
fine-tuned to no more than 10\%.\footnote{After our paper was submitted,
the authors of Ref.\cite{han} put out a revised analysis, in which they
modified their treatment of the error on the backgrounds (see footnote 4
below). Their latest analysis (arXiv:1310.4274v.3) claims a $ 3\sigma$
($5\sigma$) signal for $|\mu| = 160$ (110)~GeV at LHC14 with
3000~fb$^{-1}$. We retain mention of their earlier results in the text
to provide the reader with proper perspective for our analysis.}

Given the importance of this result, we re-examine prospects for
detection of monojet radiation
off of higgsino pair production in Sec.~\ref{sec:mjets}. Our conclusions
are, however, quite different from those of Ref.~\cite{han} since 
we find signal well below SM backgrounds (at the percent level), with no
distinctive monojet features which would allow separation of signal from
background. In Sec.~\ref{sec:mphotons}, we perform similar calculations
for mono-photon radiation and arrive at similarly pessimistic conclusions. 
We have decided such a pessimistic assessment is worthy of
publication not only because of the optimistic claims in the
literature \cite{han}, but also to highlight that claims about the
observability of monojet/mono-photon signals from  effective 
operator analyses should be viewed with caution.

\section{Prospects for monojets}
\label{sec:mjets}

To examine signal rates, we first select a low $\delew$ SUSY
benchmark model from radiatively-driven natural SUSY (RNS) 
which uses the 2-extra-parameter non-universal Higgs model (NUHM2)
with input parameters
\be
m_0,\ m_{1/2},\ A_0,\ \tan\beta,\ \mu,\ m_A
\ee
with $m_t=173.2$~GeV. 
We generate the sparticle spectrum using Isajet~7.84 \cite{isajet}. 
We fix $m_0=5$~TeV, $m_{1/2}= 750$~GeV, $A_0=-8$~TeV, $\tan\beta =10$, 
$\mu =150$~GeV and $m_A=1$~TeV. 
This leads to a sparticle spectrum with $m_{\tg}=1.9$~TeV, very heavy
squarks and sleptons, binos and winos with masses of several hundred
GeV, and a set of higgsinos with
$m_{\tw_1^\pm}=155.6$~GeV, $m_{\tz_2}=158.9$~GeV and
$m_{\tz_1}=142.2$~GeV.  These higgsinos are, of course, the focus of the
present study, and our broad conclusions are essentially independent of
the rest of the spectrum, as long as the bino and wino states are much heavier
than the higgsino states. The value of $\delew=19.7$.

We use Madgraph 5~\cite{madgraph} to generate $pp\to\tw_1^+\tw_1^-$,
$\tz_{1,2}\tz_{1,2}$ and $\tw_1^\pm\tz_{1,2}$ plus one-parton processes
(exclusive) and plus two-partons (inclusive) where for efficiency we
require the hardest final state parton to have $p_T(parton)>120$~GeV;
the final cross section is then the sum of 1-jet exclusive and 2-jet
inclusive processes.  We also evaluate the $Z+jets$, $W+jets$ and
$ZZ+jets$ backgrounds (where $Z$'s decay to neutrinos and $W$'s decay
leptonically) in a similar fashion, as the sum of one- and two-parton
processes. To avoid double-counting, we used the MLM scheme for
jet-parton matching \cite{mangano}. The events are then passed to
Pythia~\cite{pythia} for showering, hadronization and underlying
event. We have not evaluated the hard monojet background from top quark
pair production which we expect to be very small after the veto on
additional jets and leptons. This is confirmed by the results of
Ref.\cite{han}.

The Madgraph/Pythia events are then passed to the Isajet toy detector
simulation with calorimeter cell size
$\Delta\eta\times\Delta\phi=0.05\times 0.05$ and $-5<\eta<5$. The HCAL
(hadronic calorimetry) energy resolution is taken to be
$80\%/\sqrt{E}+3\%$ for $|\eta|<2.6$ and FCAL (forward calorimetry) is
$100\%/\sqrt{E}+5\%$ for $|\eta|>2.6$, where the two terms are combined
in quadrature. The ECAL (electromagnetic calorimetry) energy resolution
is assumed to be $3\%/\sqrt{E}+0.5\%$. We use the cone-type Isajet
jet-finding algorithm~\cite{isajet} to group the hadronic final states
into jets. Jets and isolated leptons are defined as follows:
\bi
\item Jets are hadronic clusters with $|\eta| < 3.0$,
$R\equiv\sqrt{\Delta\eta^2+\Delta\phi^2}\leq0.4$ and $E_T(jet)>40$~GeV.
\item Electrons and muons are considered isolated if they have $|\eta| <
  2.5$, $p_T(l)>10$~GeV with visible activity within a cone of $\Delta
  R<0.2$ about the lepton direction, $\Sigma E_T^{cells}<5$~GeV.
\item Jets with just one or three charged particles are labelled as taus.
\item  We identify hadronic clusters as
$b$-jets if they contain a B hadron with $E_T(B)>15$~GeV, $\eta(B)<3$ and
$\Delta R(B,jet)<0.5$. We assume a tagging efficiency of 60\% and
light quark and gluon jets can be mis-tagged
as a $b$-jet with a probability 1/150 for $E_{T} \leq$ 100~GeV,
1/50 for $E_{T} \geq$ 250~GeV, with a linear interpolation
for intermediate $E_{T}$ values.
\ei

Following the Atlas monojet study~\cite{atlas_monoj}, we impose the following cuts:
\bi
\item $n(jets)\le 2$ for $p_T(jet)> 30$~GeV,
\item if $n(jets) =2$, then  $p_T(j_2)<100$~GeV,
\item $b$-jet veto ($b$-jets as defined above),  to eliminate top backgrounds,
\item $\tau$-jet veto,
\item isolated lepton veto.
\ei

\begin{figure}[tbp]
\includegraphics[height=0.4\textheight]{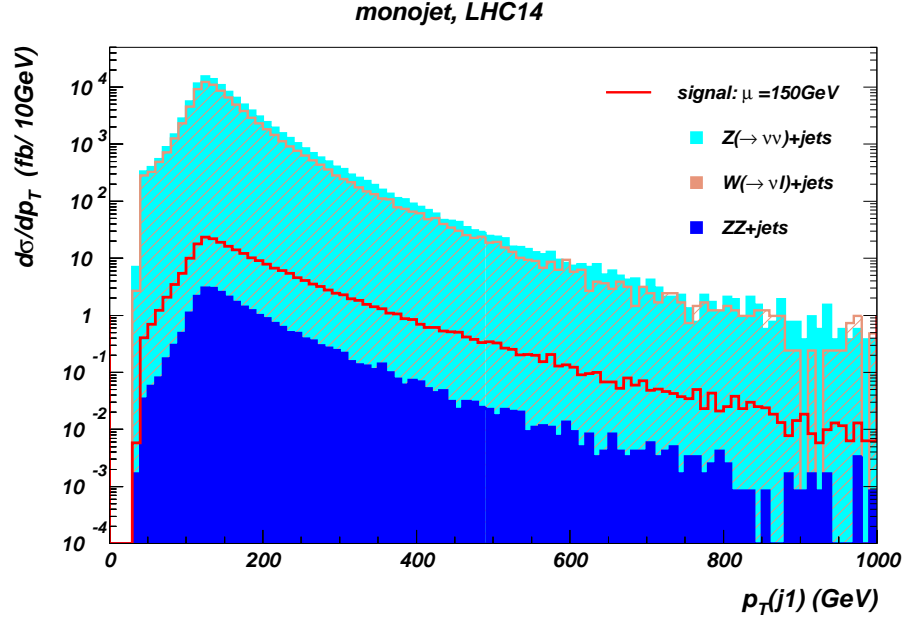}
\includegraphics[height=0.4\textheight]{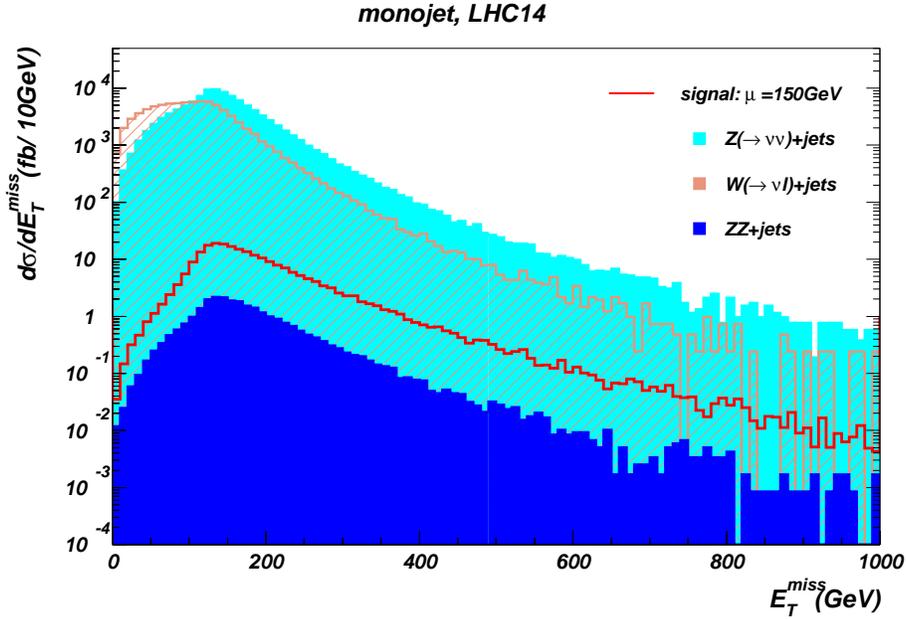}
\caption{Distribution in {\it a}) $p_T(jet)$ and {\it b}) $\eslt$ 
from initial state radiation off higgsino pair production at LHC14.
We also show backgrounds from $Z+jets$, $W+jets$ and $ZZ+jets$
production, where $W\to \ell\nu$ and $Z\to \nu\bar{\nu}$.
\label{fig:mjets}}
\end{figure}

Our resulting distributions are shown in Fig.~\ref{fig:mjets} for
{\it a}) $p_T(j_1)$ and {\it b}) $\eslt$. As expected, the shapes
of the two distributions are similar for large $p_T(j)$ and large
$\eslt$ but begin to differ for values below $\sim 200$~GeV where details of
event generation and the presence of the second jet may be important. 
From frame {\it a}), we see that $Z+jets$ production forms the dominant
background, followed closely by $W+jets$ production where the lepton
from $W$-decay is too soft or buried within a jet or too forward or otherwise
unidentified.  The signal is shown by the red solid histogram and lies
typically about two orders of magnitude below the background
distribution. We also show the distribution from $ZZ+jets$ production,
which is sub-dominant. Essentially the same qualitative
features are also seen in frame {\it b}).  Nowhere in either 
frame does the signal emerge from
background. Other cuts such as angular distributions do not help the
situation since both signal and BG are dominated by gluon radiation off
initial state quarks: really, the main difference between signal and
background as far as ISR goes is that for signal the ISR comes off a
somewhat higher $Q^2$ subprocess.  The effect of sequential cuts on the
signal and on the background is shown in Table~\ref{tab:monojet}.

%
\begin{table}
\begin{center}
\begin{tabular}{|l|r|r|r|r|}
\hline
& $Z(\nu\bar{\nu})+jets$ & $W(l\nu)+jets$ & $ZZ+jets$  & Signal \\
\hline
\hline
before cuts & 146740 & 488282 & 37.747 & 349.717 \\
\hline
$n(jets)\geq 1$ with $|\eta(j_1)|<2.0$ & 118814 & 408716 & 33.041 & 304.251 \\
$p_T(j_1)> 500$~GeV  & 816.87 & 3078.23 & 0.760 & 9.913 \\
$\eslt > 500$~GeV  & 319.83 & 380.16 & 0.370 & 6.611 \\
$\Delta \phi(j_2,\eslt) > 0.5$  & 296.54 & 300.43 & 0.333 & 5.351 \\
veto on $p_T(j_3)> 30$~GeV, $|\eta(j_3)|< 4.5$  & 249.38 & 215.36 & 0.273 & 3.544 \\
veto on e, $\mu$  & 249.38 & 45.70 & 0.273 & 2.885 \\
veto on $\tau$-jets  & 247.85 & 45.21 & 0.270 & 2.867 \\
veto on b-jets & 241.93 & 44.72 & 0.267 & 2.799 \\
\hline
\end{tabular}
\caption{Cut flow for the dominant backgrounds in the monojet 
channel and for the RNS signal with $\mu =150$~GeV. Both $Z$s in column
4 are forced to decay to neutrinos. 
All cross sections are in {\it fb}. 
\label{tab:monojet}}
\end{center}
\end{table}

Given that the signal and background have similar shapes and that $S/B
\sim 1\%$, it is very difficult to make the case that it will be
possible to realistically extract the signal \cite{giud}. Of course,
with sufficient integrated luminosity, the statistical significance will
always exceed $5\sigma$, but to claim that this means the signal is
observable means that the background is known with a precision better
than a percent!\footnote{We have traditionally included the requirement
of $S/B > 0.2$ in addition to the $5\sigma$ statistical significance
level and to a minimum 5-10 event level for the observability of the
signal. With this criterion, the signal is clearly unobservable. That
the $Z(\to \nu\nu)+jet$ background may be directly inferred from the
observed $Z(\to \ell\bar{\ell})+jet$ events does not change the
situation because the statistical fluctuations of this background remain
too large except for very high integrated luminosities. Moreover, the
theoretical systematic from the $Wj$ background still swamps the signal.
We note that up to factors of about 1.5-2, our signal and background
rates are compatible with those in Table I of Ref.~\cite{han}: {\it
i.e.}  we are in qualitative agreement with their calculation of the
cross sections.  In their analysis, Han {\it et al.}  attribute a scaled
statistical error to the $Z(\to \nu\bar{\nu})+j$ background but neglect any
systematic error on this background which can
be extracted from data, and include a statistical as well as a 10\% theoretical
uncertainty to the other backgrounds. To get the total uncertainty, they
then combine the statistical and theoretical errors in quadrature.  For
integrated luminosities of ${\cal O}(1000)$~fb$^{-1}$ that these authors
find necessary to claim a signal, the systematic error (which will not improve
with integrated luminosity), will completely dominate the
statistical error unless one assumes that the systematic uncertainties
can be reduced to about  the percent level, something we regard to be
unrealistic. Once the systematic error is properly incorporated, Han
{\it et al.}
agree with us that the signal is unobservable in the monojet
channel \cite{pvt}. }

\section{Prospects for mono-photons}
\label{sec:mphotons}

For mono-photon events (which we include for completeness), we generate the
same signal sample as before, including all higgsino pair production
reactions, but now requiring one hard photon (with $p_T > 40$~GeV)
radiation instead of a hard jet.  We also generate the background
processes $Z\gamma$ production (followed by $Z\to\nu\bar{\nu}$) and
$W\gamma$ production (followed by $W\to \ell\nu_{\ell}$ where $\ell =e$,
$\mu$ or $\tau$) as before using Madgraph plus Pythia.

For the isolated mono-photon sample, we require~\cite{atlas}:
\bi
\item $n(\gamma)\ge 1$,
\item $n(jets)\le 1$ with $|\eta (jet)|<4.5$,
\item tau-jet veto, and 
\item isolated lepton veto.  
\ei 
We regard a photon to be isolated if  
the energy in a cone of radius $\Delta R<0.4$
around photon with $p_T(\gamma )> 25$~GeV, $|\eta (\gamma )|<2.5$ is less than 5~GeV.

\begin{figure}[tbp]
\includegraphics[height=0.4\textheight]{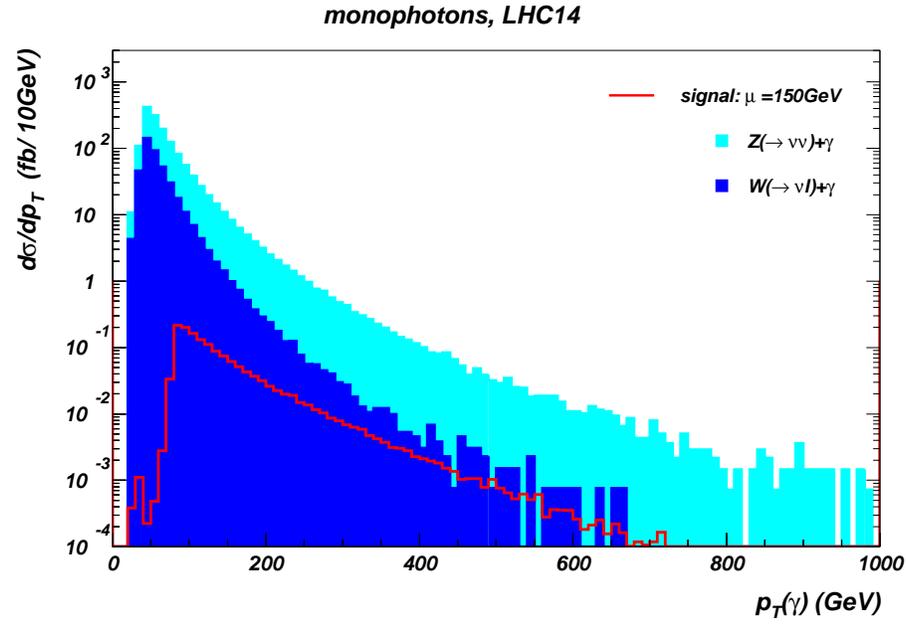}
\includegraphics[height=0.4\textheight]{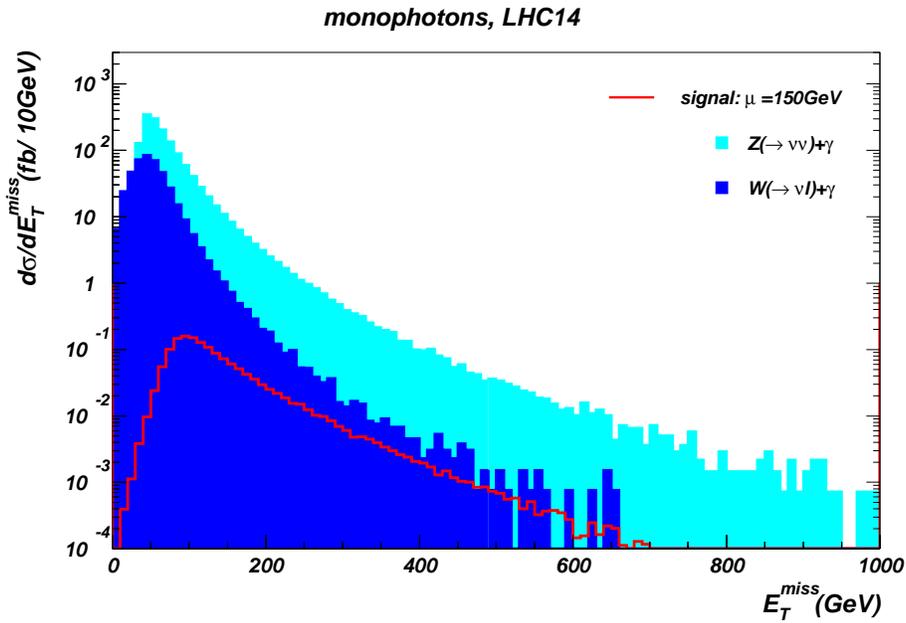}
\caption{Distribution in {\it a}) $p_T(\gamma )$ and {\it b}) $\eslt$ 
from initial state photon radiation off higgsino pair production at LHC14.
We also show backgrounds from $Z\gamma$ and  $W\gamma$ production.
\label{fig:mphotons}}
\end{figure}

Our signal and background distributions in $p_T(\gamma )$ and $\eslt$
are shown in Fig.~\ref{fig:mphotons}. As in Fig.~\ref{fig:mjets}, we see
that the shapes agree for large values of $p_T(\gamma)$ and $\eslt$.
For the entire range of $p_T(\gamma )$ as well as of $\eslt$, we again
find that signal (solid red histogram) lies below the $Z\gamma$ background by
typically two orders of magnitude. The $W\gamma$ background falls more
sharply than the $Z\gamma$ background.  This is because when we require
much higher $p_T(\gamma )$ values, then the $W$ recoils more sharply
against the gamma, and its decay products are more likely to be hard and
isolated, and to not pass the lepton/tau veto requirements.  The effect
of the sequential cuts on the signal and background cross sections is
shown in Table~\ref{tab:monophoton}. Once again, there are no distinctive
features in the distribution, and as for the monojet signal of the
previous section, we deem the mono-photon signal to be unobservable
because of the very small $S/B$ ratio.
\begin{table}
\begin{center}
\begin{tabular}{|l|r|r|r|r|}
\hline
& $Z(\nu\bar{\nu})+\gamma$ & $W(l\nu)+\gamma$ & Signal \\
\hline
\hline
before cuts & 1826.77 & 2296.76 & 3.083  \\
\hline
$n(photon)\geq 1$  & 1756.92 & 2205.32 & 2.895 \\
$p_T(\gamma 1)> 150$~GeV  & 63.79 & 75.58 & 0.997 \\
$\eslt > 150$~GeV  & 49.43 & 15.42 & 0.778 \\
$n(jets)\leq 1$, $|\eta(jet)|< 4.5$  & 39.60 & 9.43 & 0.473 \\
veto on e, $\mu$  & 39.60  & 2.67 & 0.418  \\
veto on $\tau$-jets  & 36.15 & 2.45 & 0.371 \\
\hline
\end{tabular}
\caption{Cut flow for the dominant backgrounds in the monophoton  
channel and for the RNS signal with $\mu =150$~GeV. 
All cross sections are in {\it fb}. 
\label{tab:monophoton}}
\end{center}
\end{table}

\section{Concluding remarks} 

The existence of light higgsinos with masses smaller than 200-300~GeV
(depending on how much fine-tuning one is willing to tolerate) is a
robust feature of natural SUSY models. Although these higgsinos can 
be pair produced at large rates at the LHC, 
the signal will be buried below SM backgrounds because of the 
small energy release from their decays. 
In this paper, we have examined prospects for their detection  via 
pair production in association with a hard jet or a hard, isolated photon
resulting in characteristic monojet or mono-photon events at LHC14.
We emphasize here that constraints obtained from analyses \cite{cms,atlas}
using contact interactions between quarks and the higgsinos are
inapplicable in this connection because the effective operator approximation
fails badly for higgsino pair production.

While monojet and mono-photon signal events indeed occur at an
observable rate particularly at the luminosity upgrade of the LHC, we
are pessimistic about the prospects for their detection because
backgrounds from $Z$ and $W$ production in association with a jet or an
isolated photon overwhelm the signal by two orders of magnitude even for
very large values of jet or photon transverse momentum and $\eslt$ in
these events. It seems to us difficult to imagine that it would be
possible to claim a signal for new physics 
in these channels based solely on an excess
of ${\cal O}(1\%)$ without an observable distortion of any distribution.


In arriving at our negative conclusion, we should mention that have not
investigated whether it might be possible to extract the higgsino signal
by examining the soft debris from the decays of $\tw_1$ and $\tz_2$
produced via $\tw_1\tz_2$, $\tz_1\tz_2$ and $\tw_1\tw_1$ pair production
processes that dominate higgsino pair production~\cite{rnslhc}. This
will require a careful analysis of potential backgrounds from higher
order Standard Model processes. Despite our cautious pessimism, we leave
open the possibility that a clever analysis may make it feasible to
tease out this signal at a luminosity upgrade of LHC14.

\section*{Acknowledgments}
A.M. thanks CERN Theory Division for its support and hospitality. 
This work was supported in part by the US Department of Energy, Office of High
Energy Physics.

\section*{Note Added} After this study was completed we saw Ref.~\cite{sz}
  which claims that exclusion (not discovery) of electroweak-ino masses
  up to 200~GeV
is possible with 300~fb$^{-1}$ of integrated luminosity at LHC14 if the
systematic error can be reduced to the 1\% level.

%

%

\begin{thebibliography}{99}
\small





\bibitem{wss} See, {\it e.g.} H.~Baer and X.~Tata, {\it Weak Scale Supersymmetry: From
Superfields to Scattering Events},
(Cambridge University Press, 2006).
%
\bibitem{ltr}  H.~Baer, V.~Barger, P.~Huang, A.~Mustafayev and X.~Tata,
  Phys.\ Rev.\ Lett.\  {\bf 109} (2012) 161802.
%
\bibitem{rns} H.~Baer, V.~Barger, P.~Huang, D.~Mickelson, A.~Mustafayev
and X.~Tata, 
\prd{87}{2013}{035017} and \prd{87}{2013}{115028}.
%
\bibitem{ellis} J.~Ellis, K.~Enqvist, D.~Nanopoulos and F.~Zwirner,
  Mod. Phys. Lett. {\bf A1} (1986) 57.
%
\bibitem{bg} R.~Barbieri and G.~Guidice, \npb{306}{1988}{36}. 
%
\bibitem{kn} R. Kitano and Y. Nomura, \plb{631}{2005}{58}, 
   \prd{73}{2006}{095004} and \hepph{0606134}.
%
\bibitem{papucci} M.~Papucci, J.~T.~Ruderman and A.~Weiler,
  \jhep{1209}{2012}{035}.
%
\bibitem{comp} H.~Baer, V.~Barger and D.~Mickelson, \prd
  {88}{2013}{095013}.
%
\bibitem{sundrum} C.~Brust, A.~Katz, S.~Lawrence and R.~Sundrum,
  \jhep{1203}{2012}{103}.
%
\bibitem{atlas_h} G.~Aad {\it et al.}  [ATLAS Collaboration], 
\plb{716}{2012}{1}.
%
\bibitem{cms_h} S.~Chatrchyan {\it et al.}  [CMS Collaboration],  
\plb{716}{2012}{30}.
%
%
\bibitem{atlas_susy}
  G.~Aad {\it et al.}  [ATLAS Collaboration],
 \prd{87}{2013}{012008}; ATLAS-CONF-2013-04 and ATLAS-CONF-2013-061.
%
\bibitem{cms_susy}
  S.~Chatrchyan {\it et al.}  [CMS Collaboration],
  \jhep{1210}{2012}{018} and CMS-PAS-SUS-13-007.
%
\bibitem{lhcltr} H.~Baer, V.~Barger, P.~Huang, D.~Mickelson, A.~Mustafayev, W.~Sreethawong and X.~Tata,
  Phys.\ Rev.\ Lett.\  {\bf 110} (2013) 151801.
%
\bibitem{rnslhc} H.~Baer, V.~Barger, P.~Huang, D.~Mickelson, A.~Mustafayev, W.~Sreethawong and X.~Tata,
  arXiv:1310.4858 [hep-ph].
%
\bibitem{matchev} S.~Arrenberg, H.~Baer, V.~Barger, L.~Baudis, D.~Bauer, J.~Buckley, M.~Cahill-Rowley and R.~Cotta {\it et al.},
  arXiv:1310.8621 [hep-ph].
%
\bibitem{zhang} H.~Zhang, Q.~Cao, C.~Chen and C. Li, JHEP 1108 (2011)
  018.
%
\bibitem{beltram} M.~Beltram {\it et al.} JHEP 1009 (2010) 037.
%
\bibitem{goodman} J.~Goodman {\it et al.} \prd {82}{2010}{116010}.
%
\bibitem{rajaraman} A.~Rajaraman, W.~Shepherd, T.~Tait and A.~Wijangco,
  Phys. Rev. {\bf D84} (2011) 095013.
%
\bibitem{fox} P.~Fox, R.~Harnik and J.~Kopp, Phys. Rev. {\bf D85} (2012)
056011.
%
\bibitem{cms} S.~Chatrchyan {\it et al.} (CMS Collaboration) JHEP, 1209
  (2012) 094 (jets) and Phys. Rev. Lett. {\bf 108} (2012) 261803.
%
\bibitem{atlas} G.~Aad {\it et al.} (ATLAS Collaboration) JHEP 1304
  (2013) 075 (jets) and Phys. Rev. Lett. {\bf 110} (2013) 011802
  (photons).
%
\bibitem{buchmueller} O.~Buchmueller, M.~Dolan and C.~McCabe,
  arXiv:1308:6799. 
%
\bibitem{han} C.~Han, A.~Kobakhidze, N.~Liu, A.~Saavedra, L.~Wu and J.~M.~Yang,
  arXiv:1310.4274 [hep-ph].
%
%
\bibitem{isajet} ISAJET, by H.~Baer, F.~Paige, S.~Protopopescu and
X.~Tata, \hepph{0312045}.
%
\bibitem{madgraph} J.~Alwall, M.~Herquet, F.~Maltoni, O.~Mattelaer and T.~Stelzer,
  \jhep{1106}{2011}{128}.
%
\bibitem{mangano} M.~Mangano, M.~Moretti, F.~Piccinini and M.~Treccani, \jhep{0701}{2007}{013}.

%
\bibitem{pythia} T.~Sjostrand, S.~Mrenna and P.~Z.~Skands,
  \jhep{0605}{2006}{026}.
%
\bibitem{atlas_monoj} Atlas collaboration, ATLAS-CONF-2011-096.
%
%
\bibitem{giud} G.~Giudice, T.~Han, K.~Wang and L.-T.~Wang,
  Phys. Rev. {\bf D81} (2010) 115011.
%
\bibitem{pvt} L. Wu and J.~M.~Yang, private communication. 
%
\bibitem{sz} P.~Schwaller and J.~Zurita, arXiv:1312.7350 (2013).
%
\end{thebibliography}
\end{document}